\documentclass[preprint,prb,superscriptaddress,showpacs]{revtex4-1}
\usepackage{bm}
\usepackage{color}
\usepackage{graphicx}
\usepackage{amssymb}
\usepackage{amsmath}
\usepackage{amsfonts}
\usepackage{ulem}
\usepackage{rotating}
\usepackage{multirow}
\def\urs{URu$_2$Si$_2$}

\begin{document}

\title{Study of the magnetic properties of \urs\ under uniaxial stress by neutron scattering.}

\author{Frederic Bourdarot}
\affiliation{SPSMS, UMR-E 9001, CEA-INAC/ UJF-Grenoble 1, MDN, 17 rue des martyrs, 38054 Grenoble, France}

\author{Nicolas Martin}
\affiliation{SPSMS, UMR-E 9001, CEA-INAC/ UJF-Grenoble 1, MDN, 17 rue des martyrs, 38054 Grenoble, France}

\author{Stephane Raymond}
\affiliation{SPSMS, UMR-E 9001, CEA-INAC/ UJF-Grenoble 1, MDN, 17 rue des martyrs, 38054 Grenoble, France}

\author{Louis-Pierre Regnault}
\affiliation{SPSMS, UMR-E 9001, CEA-INAC/ UJF-Grenoble 1, MDN, 17 rue des martyrs, 38054 Grenoble, France}

\author{Dai Aoki}
\affiliation{SPSMS, UMR-E 9001, CEA-INAC/ UJF-Grenoble 1, IMAPEC, 17 rue des martyrs, 38054 Grenoble, France}

\author{Valentin Taufour}
\affiliation{SPSMS, UMR-E 9001, CEA-INAC/ UJF-Grenoble 1, IMAPEC, 17 rue des martyrs, 38054 Grenoble, France}

\author{Jacques Flouquet}
\affiliation{SPSMS, UMR-E 9001, CEA-INAC/ UJF-Grenoble 1, IMAPEC, 17 rue des martyrs, 38054 Grenoble, France}

\date{\today}

\begin{abstract}
The aim of this study is to compare the magnetic behavior of \urs\ under uniaxial stress along the \textbf{a}-axis with the behavior under hydrostatic pressure. Both are very similar, but uniaxial stress presents a critical stress $\sigma_x^a$ smaller (0.33(5)GPa) than the hydrostatic critical pressure p$_x$ =0.5 GPa where the ground state switches from HO (hidden order) to AF (antiferromagnetic) ground state. From these critical values and from Larmor neutron diffraction (LND), we conclude that the magnetic properties are governed by the shortest U-U distance in the plane (\textbf{a} lattice  parameter). Under stress, the orthorhombic unit cell stays centered. A key point shown by this study is the presence of a threshold  for the uniaxial stress along the \textbf{a}-axis before the appearance of the large AF moment which indicates no-mixture of order parameter (OP) between the HO ground state and the AF one as under hydrostatic pressure. The two most intense longitudinal magnetic excitations at \textbf{Q$_0$}=(1,0,0) and \textbf{Q$_1$}=(0.6,0,0) were measured in the HO state: the excitation at \textbf{Q$_0$} decreases in energy while the excitation at \textbf{Q$_1$}  increases in energy with the uniaxial stress along the \textbf{a}-axis. The decrease of the energy of the excitation at \textbf{Q$_0$} seems to indicate a critical energy gap value of 1.2(1) meV at $\sigma_x^a$.  A similar value was derived from studies under hydrostatic pressure at p$_x$.

\end{abstract}

\pacs{75.30.Mb, 75.50.Ee, 25.40.Dn, 78.70.Nx, 61.50.Ks, 75.25.Dk, 75.30.Kz}

\maketitle

\section{Introduction}
\begin{figure}
\includegraphics[width=76mm]{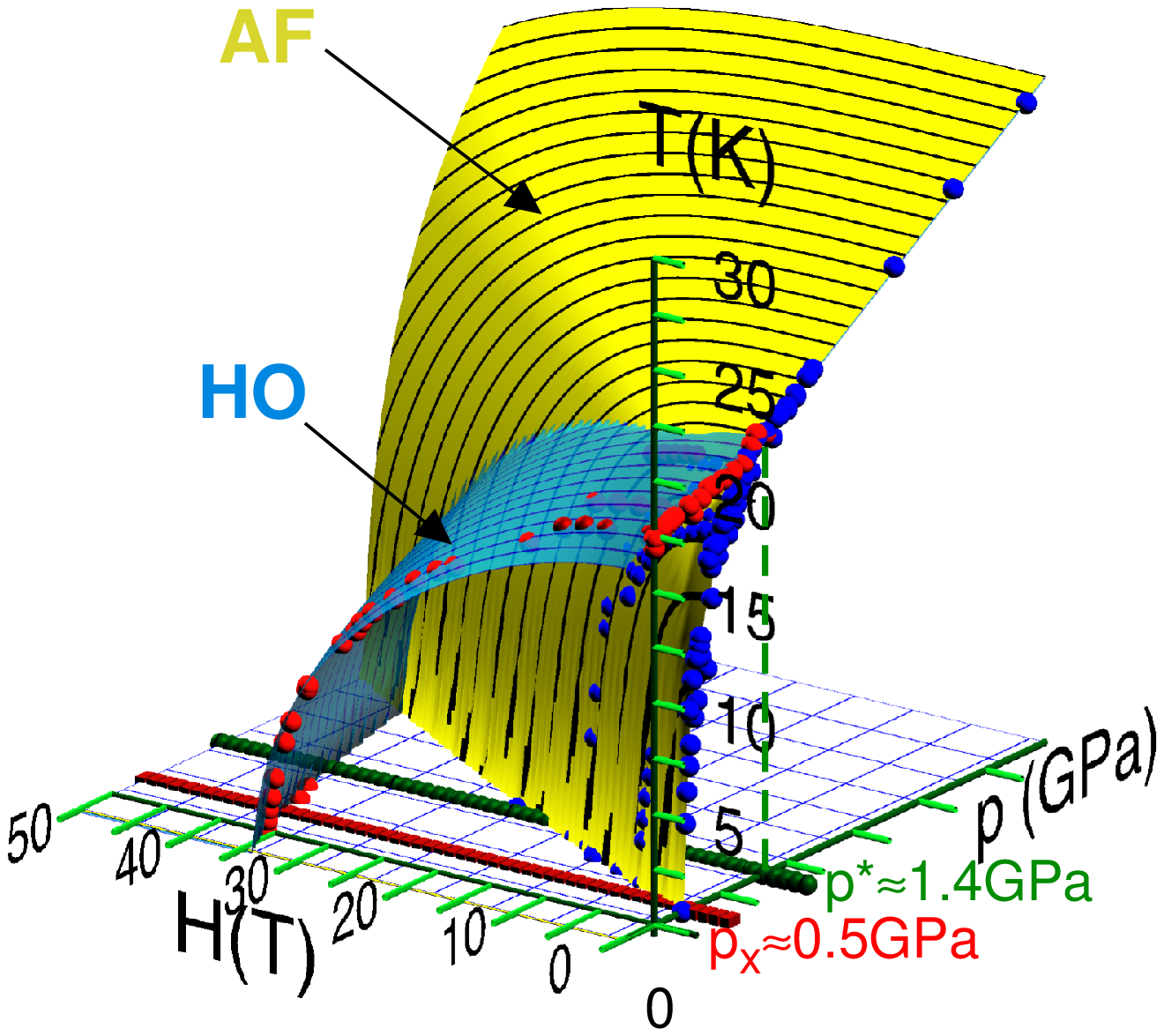}
\caption{Schematic phase diagram $(T,H,p)$ of \urs. The data come from Ref.[\onlinecite{Hassinger:2008,Aoki:2009,Aoki:2010,Jaime:2002}]. p$_x\simeq0.5$ GPa corresponds to the critical pressure, p$^*\simeq1.4$ to the pressure where \urs\ transits directly from PM state to AF state. The superconducting phase is not presented to simplify the phase diagram.}
\label{fig0}
\end{figure}

Puzzling heavy fermion physicists for more than twenty years, \urs\ is one of the most studied and least understood uranium compounds. The mysterious phase transition at T$_0 \sim$ 17.8 K of this $5f$ heavy-electron compound is characterized by large bulk anomalies and sharp magnetic excitations in q-space and energy, at different \textbf{Q}-vectors (\textbf{Q$_0$}=(1,0,0) and \textbf{Q$_1$}=(0.6,0,0)). Concomitant with this order, a tiny but persistent antiferromagnetic moment ($\sim$ 0.02 $\mu_B$) is measured in all samples with a wave-vector \textbf{Q$_{AF}$}=(0,0,1) (equivalent to \textbf{Q$_0$}). This tiny staggered moment is difficult to consider as the order parameter in a conventional antiferromagnetism frame as it cannot be reconciled with a jump of the specific heat $\Delta C/T \sim$ 300 mJ/K$^2$mol involving an entropy (S $\sim$ 0.2R$\ln$(2)). Because there is no determination of the order parameter (OP), the order in \urs\ is named the hidden order (HO). However, under pressure, \urs\ orders in a high moment antiferromagnetic (AF) structure with the wave-vector \textbf{Q$_{AF}$} and a moment of 0.36-0.4$ \mu_B$\cite{Bourdarot:2004b,Bourdarot:2005}. The well-defined phase diagram (see Fig.\ref{fig0}) shows that when \urs\ switches from HO to AF state at a critical pressure p$_x\simeq$ 0.5 GPa, the bulk superconductivity disappears \cite{Hassinger:2008,Hassinger:2008a} as well the antiferromagnetic excitation E$_0$ at \textbf{Q$_0$}, signature of the HO phase \cite{Villaume:2008}. At p$_x$, the excitation E$_1$ at \textbf{Q$_1$} jumps from 5 meV to 8 meV\cite{Villaume:2008}. Under magnetic field (applied along \textbf{c}-axis), the pressurized AF phase is unstable and \urs\ reenters into the HO state\cite{Aoki:2009,Aoki:2010}(see Fig.\ref{fig0}). 

Thermal expansion measurements established that stress will increase T$_0$ when it is applied along the \mbox{\textbf{a}-axis} and decrease T$_0$ when it is applied along the \mbox{\textbf{c}-axis}. The opposite effect is observed for the evolution of superconductivity\cite{Bakker:1992,Guillaume:1999}. Recently a theoretical model \cite{Harima:2010} indicated that the space group of the PM (paramagnetic) state and of the HO state are different but may keep the atomic positions which explains why most of the local probes did not detect any modification in the crystal structure. This new point of view led us to study the variation of the crystallographic structure and revisit the magnetic behavior of \urs\ under uniaxial stress along the \textbf{a}-axis. In a first experiment, Yokoyama \textit{et al.} have shown that under uniaxial stress along the \textbf{a}-axis and below T$_0$ the HO state is gradually mixed to the AF state\cite{Yokoyama:2005b}. 

Uniaxial stress combined with hydrostatic pressure can bring information about the exchange integrals between magnetic atoms. For example when an AF compound switches to PM state under pressure, a study under uniaxial stress can determine which parameter (lattice parameter or ratio of lattice parameters) governs the magnetic behavior of the system. However uniaxial stress is a tool which is not used very often with neutron scattering as there is a high probability to break the sample and as large crystals with good ratio of the height by the diameter must be selected to realize a homogeneous uniaxial stress conditions. In spite of this, neutron users prefer to use thin samples as  for the first experiment on \urs\cite{Yokoyama:2005b} with the difficulty of poor homogeneity.

This paper is organized as follows. In section \ref{II}, we present the experimental set-up. The results are presented in section \ref{III}. Section \ref{IV} is dedicated to the discussion of our results and their comparisons with previous data. Finally concluding remarks are given in section \ref{V}.

\section{Experimental set-up}
\label{II}
Uniaxial stress was applied along the \textbf{a}-axis of two single crystals of \urs\  coming from different batches. The first one is a small sample with a perfect cylindric shape of diameter (d) 3.78 mm and height (h) 1.66 mm  and the second one has a parallelepiped shape of surface 12 mm$^2$ with a vertical \textbf{a}-axis of 8 mm length. An important parameter to perform reliable uniaxial stress experiments is, as we will see latter, the ratio of the height by the diameter of the sample that we define as $\kappa=h/d$: $\kappa \simeq$ 0.5 for the small sample and $\kappa \simeq$ 2.0 for the large one (In this case an average diameter was calculated). The small sample comes from the same batch as samples  previously used for high field \cite{Bourdarot:2003} and  pressure measurements \cite{Villaume:2008}. The second sample is from a new crystal grown in a tetra-arc furnace and annealed for 5 days at 1075$^\circ$C. The samples were installed successively at the bottom of the Institut-Laue-Langevin (ILL) uniaxial stress stick between two foils of cadmium and gold to flatten the surface defects of the sample and loading platforms,  and  to reduce the friction when the sample is pressed. The stick was installed in an ILL-orange cryostat on the cold-triple-axis IN12, CRG spectrometer at ILL.

The small sample was principally used to determine the nuclear structure with neutrons at k$_f$=1.48\AA$^{-1}$. Monochromator and analyzer were put flat vertically and horizontally, in order to enhance the effect of the graphite filter. The collimators were: open-$60^{\prime}$-$60^{\prime}$-$60^{\prime}$. Two filters: one Be filter on k$_i$ and one graphite filter on k$_f$ (the graphite filter was oriented such that the neutrons of wavelength $\lambda$/2 were diffracted out by the reflection (006) of the graphite; the neutrons of wavelength $\lambda$ are not diffracted out as the reflection (003) of the graphite does not exist). The large sample was used for inelastic and elastic magnetic scattering measurements with k$_f$ =1.5\AA$^{-1}$. Vertically curved monochromator and horizontally curved analyzer were used with collimators: open-$60^{\prime}$-open-open, one Be filter on k$_f$, plus a not-cooled Be filter which was installed on k$_i$ when magnetic Bragg peaks were measured.

\begin{figure}
\begin{center}
	\includegraphics[width=.5\textwidth]{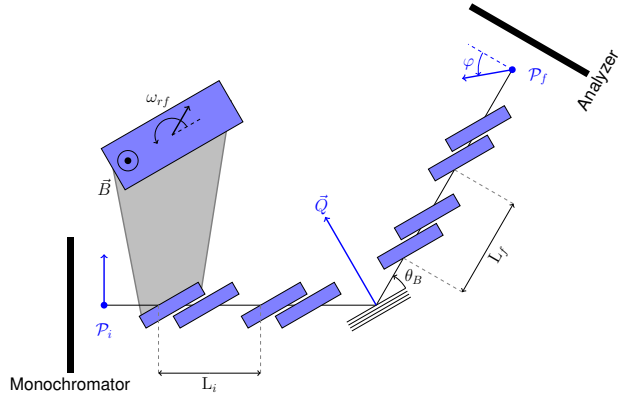}
\end{center}
\caption{Sketch of a Neutron Larmor Diffraction setup. The scattering plane is horizontal. Radio frequency spin flippers (in blue) are rotated so that their faces are parallel to the lattice planes which generate the process of diffraction. This peculiar arrangement cancels out the effect of beam divergence in the scattering plane.}
\label{fig:NLD_1}
\end{figure}

A complementary Neutron Larmor diffraction (NLD) experiment was performed on IN22 (CEA CRG-beam-line at ILL) on the large sample to obtain the distribution and the temperature dependence of the lattice parameters \textbf{a} and \textbf{c}. NLD exploits the Larmor precession of the neutron spin within well-defined magnetic field regions to measure the particle's wavelength with high accuracy ({\it Larmor encoding}). The beam is initially polarized by reflection on a Heusler-111 monochromator. The field is in practice simulated by pairs of radio frequency spin flippers (RFSFs) separated by a magnetically screened volume (Bootstrap technique described in Ref.[\onlinecite{Gahler:1988}]). These devices consist of a rectangular coil producing a vertical static field $\vec{B_{0}}$ and containing another coil generating a smaller field  oscillating in the horizontal plane at a frequency $\omega_{rf} = \gamma_{n}B_{0}$ ($\gamma_{n}$ is the neutron gyromagnetic ratio). The length of the precession region is defined as four times the distance (bootstrap set-up) between the first and second RFSF in each arm of the spectrometer, respectively $L_{i}$ and $L_{f}$ (see Fig.\ref{fig:NLD_1}). Larmor encoding is used to perform high resolution diffraction measurement thanks to the symmetry of Bragg's law which selects the neutron's wavelength for a particular lattice spacing $d$. Consequently, the total phase, namely the rotation angle of the magnetic moment of the neutron, is defined as $\varphi = \omega_{rf}\ 4\  (L_{i} + L_{f}) \ m_{n}  / h \cdot d\cdot \sin(\theta_B)$ and measured by projection on the quantization axis of a Heusler analyzer ($m_{n}$ is the neutron mass and $h$ is the Planck's constant). Rotating the RFSFs so that their faces are parallel to the lattice planes generating diffraction ensures that the latter relation is fulfilled for any wavelength in the incoming beam's bandwidth. One finally obtains the very simple equality $\delta \varphi / \varphi_{0} = \delta d / d_{0}$ where subscripted variables are reference quantities. The advantage of this method is that the high resolution is achieved with modest beam collimation as a d-spacing creates the same phase for all neutrons. The experiment was performed with an incident wavelength k$_i$=2.662 \AA$^{-1}$. Using a Larmor frequency $\omega_{rf}$ =  670 kHz and a total effective length ($4(L_i+L_f)$) of 3.44 m, the total Larmor phases $\varphi$ was close to 8650 rad with a theoretical resolution of the order of $5\cdot10^{-6}$. The technical configuration is described in Ref.[\onlinecite{Martin:2011}].

\section{Results}
\label{III}
\subsection{Elastic nuclear scattering}
\label{IIIA}
\urs\ crystallizes in the body-centered-tetragonal structure (space group: $I4/mmm$). Only 3 parameters are needed to describe the nuclear cell: \textbf{a} and \textbf{c}, the lattice constants and \textit{z$_{Si}$}, the silicon atomic position along the \textbf{c}-axis. It is now believed \cite{Harima:2010} that the symmetry goes from body-centered-tetragonal in the paramagnetic state to simple-tetragonal in HO. Both space groups have the particularity to keep the same atomic positions so to conserve the same nuclear structure factor in both states. As uniaxial stress is applied along one \textbf{a}-axis (named \textbf{a$_v$}, \textit{v} for vertical), the tetragonal symmetry is broken and the second \textbf{a}-axis (named \textbf{a$_h$}, \textit{h} for horizontal) is no longer equivalent to \textbf{a$_v$}: the symmetry becomes orthorhombic. As measurements are performed on a triple-axis spectrometer only few nuclear reflections are available in the plane (\textbf{a$_h$},\textbf{c}). However, the nuclear reflections \textbf{Q$_0$}=(1,0,0) and \textbf{Q}=(0,0,1) stay forbidden under uniaxial stress in the paramagnetic state (above T$_0$) which indicates that the translation (1/2, 1/2, 1/2) is conserved in the crystal. To confirm this assumption, it must be checked that the nuclear reflection \textbf{Q$_0$}=(0,1,0) stays forbidden. This means that only the 4-fold axis disappears and the space group becomes \textit{I $mmm$} in the paramagnetic state under uniaxial stress. In this new space group the nuclear structure is described with a new parameter: the position along \mbox{\textbf{c}-axis} of the ruthenium atom (z$ _{Ru}$).

%It is interesting to note that, under uniaxial stress along the \textbf{a}-axis ($\sigma^a$), the space group of the PM state (\textit{I$_{mmm}$}) may also be transformed in the HO state into a space group with the same atomic positions. The space group of the HO state would then be $Pnnm$ or $Pnnn$. This may indicate an important property of the U site symmetry: the transition into HO state does not change the atomic position and HO state exists even if the four-fold axis is broken to become a two-fold axis.

It is interesting to note that, as the space group $I4/mmm$, the space group of the PM state ($I mmm$) under uniaxial stress along the \textbf{a}-axis ($\sigma^a$) may also be transformed in the HO state into a space group with the same atomic positions. The space group of the HO state would then be $Pnnm$ or $Pnnn$. This may indicate an important property of the U site symmetry: the transition into HO state does not change the atomic position and HO state exists even if the four-fold axis is broken to become a two-fold axis.

The pressure dependence of the lattice parameters were determined at 8K on the small sample using the positions of the nuclear Bragg reflections \textbf{Q}=(0,0,2) and \textbf{Q}=(1,0,1). Both lattice parameters in the scattering plane increase slightly and almost linearly with the pressure as expected in the elastic deformation regime. As the increase of $\partial c/\partial \sigma^a$ is approximatively 2.5 larger than $\partial a_h/\partial \sigma^a$ which is almost the ratio $c/a$, the ratio $c/a_h$ remains almost constant under uniaxial stress for the range of pressure from 0.2 to 0.65 GPa.  Compared with the dependences of the cell parameters deduced from the elastic constants (given in Ref.[\onlinecite{Wolf:1994}]), there is a quite good agreement. 

The intensities of the nuclear reflections at \textbf{Q}=(0,0,2) and at \textbf{Q}=(1,0,1) increase by 77\% and 88\% respectively at the maximum of the applied uniaxial stress. Even if these reflections have a small structure factor compared to the nuclear structure factor of the largest intense reflection \textbf{Q}=(2,0,0) (1/5 and 1/12 respectively), they are largely affected by the extinction because of the size of the sample and the long wavelength of the neutrons. However the larger increase of the weaker reflection indicates a modification of the nuclear structure factor certainly due to a modification of $z_{Si}$. This effect will be studied in a future dedicated diffraction experiment.

\subsection{Elastic magnetic scattering}
\label{IIIB}
\begin{figure} %
\includegraphics[width=80mm]{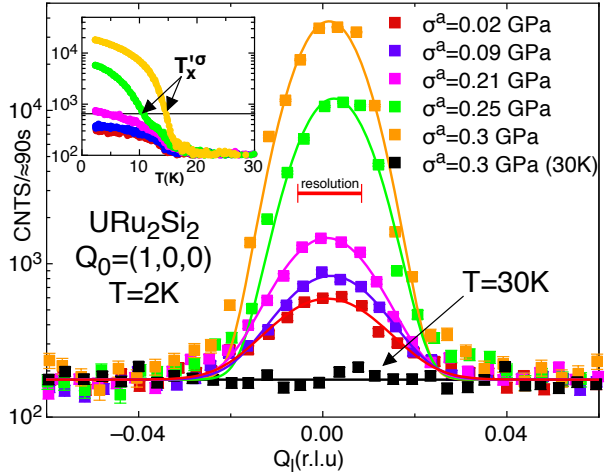}
\caption{Intensity in log scale of the magnetic Bragg peak at \textbf{Q$_0$} for different uniaxial stress at low temperature T=4K and above T$_0$(T=30K) for the largest stress. The inset displays the temperature variation of the magnetic intensity (also in a log scale) and shows the different regime according \urs\ is in the HO or in the AF states. T$^{'\sigma}_x$ is defined in the text.}
\label{omegab}
\end{figure}

\begin{figure} %
\includegraphics[width=80mm]{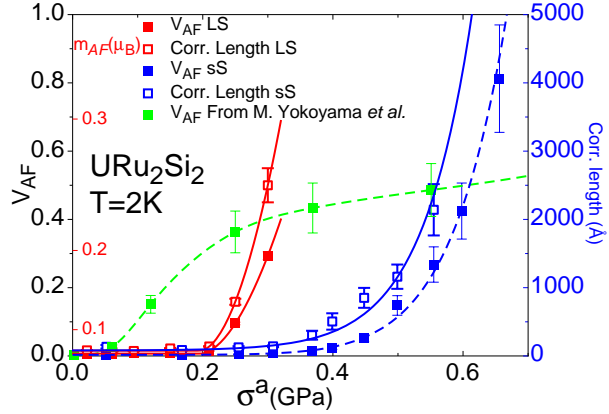}
\caption{Uniaxial stress dependence of the AF volume and of the correlation length for our two crystals and also the AF volume of the crystal determined by M. Yokoyama in Ref.[\onlinecite{Yokoyama:2005b}]. The AF volume is determined with the assumption that the moment in the AF state is m$_{AF}$=0.36$\mu_B$. For indication, the correspondence between AF volume and an intrinsic moment value is given, but noticed that this scale (in red) is not linear. LS and sS mean large and small samples respectively.}
\label{AFM}
\end{figure}

%The magnetic Bragg peaks at \textbf{Q$_0$} measured on the large sample at 30 K, have no intensity for the whole range of stress above T$_0$. 
The magnetic Bragg peak measured at \textbf{Q$_0$} and at 30 K, on the large sample, presents no intensity for the whole range of measured stress. Transverse scans are presented in Fig.\ref{omegab} for different stress at low temperature and at T=30K for the largest stress we applied: $\sigma^a$=0.3 GPa. The resolution of the spectrometer of 9.12(2)$\cdot$10$^{-3}$\AA$^{-1}$ was determined using the widths of the two nuclear peaks at \textbf{Q}=(1,0,1) and \textbf{Q}=(0,0,2). To determine the magnetic correlation lengths, the magnetic peaks were fitted as a convolution of the gaussian resolution with a lorentzian function. The magnetic intensities given by neutron scattering at ambient pressure can be associated to the AF volume fraction assuming that it comes from residual AF components induced by local defects inducing local stresses. At low temperature (T=2K), the magnetic intensity increases slightly for stresses lower than 0.2 GPa and drastically for larger stresses. The stress dependence of the AF volume, assuming a saturated AF moment in the AF state of 0.36$\pm$0.04  $\mu_B$ (mean value of AF moment of \urs\ in the AF phase under hydrostatic pressure), and the magnetic correlation length at low temperature are presented in Fig.\ref{AFM}. Both start to diverge at $\sigma^a=0.3$ GPa and we estimate the critical stress $\sigma_x^a \approx$ 0.33GPa which is defined as half of the crystal volume in the AF state. With the hypothesis of a linear relation between magnetic intensity and AF volume, the repartition of volumes is 30\%/70\% of AF/HO states at $\sigma^a$=0.3 GPa. A critical temperature T$_x^{\sigma}$ can be defined as well when half of the crystal volume in the AF state: T$_x^{\sigma}$ will arise from 0 K just above $\sigma_x^a$. However, this estimation is based on the assumption of identical AF moment values when uniaxial or hydrostatic pressure is applied, which may be wrong as the tetragonal symmetry is broken with the stress.

The temperature dependence of the AF moment  presents different behaviors according to whether the state is HO  or AF. This modification happens at a value of the AF moment around 0.06 $\mu_B$ and can be chosen as a criterion for the transition temperature T$_x^{'\sigma}$  from HO to AF order (see inset figure \ref{omegab}). T$_x^{\sigma}$ is larger than T$_x^{'\sigma}$ as it corresponds to a magnetic intensity equivalent to 0.25 $\mu_B$. Similar behavior was found in previous measurements under hydrostatic pressure \cite{Bourdarot:2004b,Yokoyama:2005b,Butch:2010}.

A larger critical stress $\approx$ 0.55 GPa was determined for the small sample. This difference is due to the reduced free expansion condition when the ratio $\kappa$ is small. It is well-known that for samples with small $\kappa$ ratio ($\kappa<2$) as for our small crystal or for samples of M. Yokoyama \textit{et al.}\cite{Yokoyama:2005b} ($\kappa\simeq0.2$), the experimental conditions are between uniaxial stress (free lateral expansion) and uniaxial strain (no free lateral expansion) which, in the last condition, increases the critical pressure. The behavior of the AF volume versus uniaxial stress in the sample of the Ref.[\onlinecite{Yokoyama:2005b}] (plotted to comparison in Fig \ref{AFM}) cannot  only be explained by the experimental conditions. The increase of the AF volume is largely affected by the quality of the sample: it shows a large AF volume already at low hydrostatic pressure\cite{Amitsuka:1999} which does not correspond to the usual results\cite{Amitsuka:2008}.

\subsection{Inelastic magnetic scattering at Q$_0$ and Q$_{inc}$}
\label{IIIC}
\begin{figure} %
\includegraphics[width=80mm]{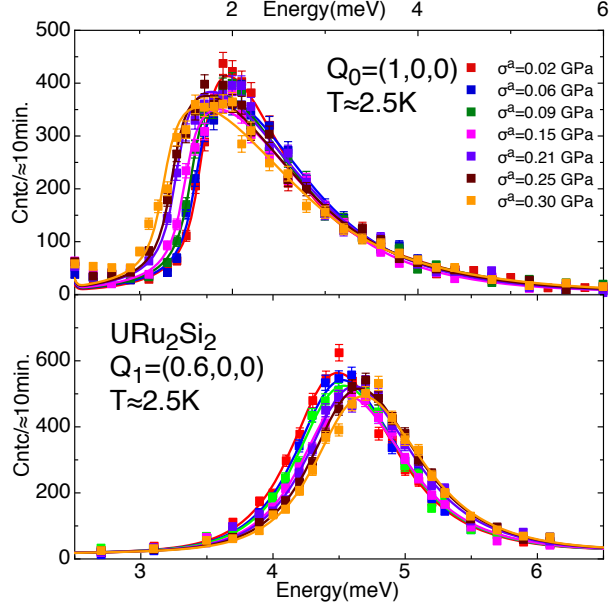}
\caption{Magnetic excitations at \textbf{Q$_0$}=(1,0,0) and \textbf{Q$_1$}=(0.6,0,0) for $\sigma^a$ from 0 to 0.3 GPa at low temperature.}
\label{E0E1b}
\end{figure}

For the inelastic neutron scattering measurements at \textbf{Q$_0$}, the variation with stress of the intensity of the excitation is small (see Fig.\ref{E0E1b}) and the energy gap E$_0$ decreases linearly from 1.68(1)meV down to 1.27(1)meV. To evaluate the variation of this intensity some assumptions were made. First, we consider that the intensity of the excitation is related to the volume of the HO phase, which means that this excitation is characteristic of the HO. Secondly, \urs\ can be described by a single-mode approximation: sharp dispersion at the antiferromagnetic position \textbf{Q$_0$}, which gives the intensity of the magnetic excitation $\sim 1/E_0$ \cite{Zaliznyak:1998}. Then the variation of the intensity as a function of uniaxial stress should be I($\sigma^a$)=I($\sigma^a$=0)V$_{HO}$$E_0$($\sigma^a$=0)/$E_0$($\sigma^a$). This gives only a slight decrease of the intensity in agreement with the inelastic neutron scattering measurements. There is no large modification of the intensity because at the larger stress we applied most of the sample was still in the HO state. The evolution of the energy gap and of the intensity versus uniaxial stress along \textbf{a}-axis are very similar to the results under hydrostatic pressure \cite{bourdarot:2010}. Considering the critical stress $\sigma^a_x$=0.33 GPa, determined from the elastic magnetic measurements, we can deduce a critical energy gap E$_{0-\mathrm{crit}}$ of 1.2 meV at $\sigma^a_x$ which has the same value  than the critical energy gap under hydrostatic pressure at p$_x$. However the critical pressure p$_x$ is two times larger (0.5-0.7 GPa\cite{Villaume:2008,Amitsuka:2008}) than the critical stress $\sigma^a_x$.

For the incommensurate excitation at \textbf{Q$_1$}, the energy gap slightly increases with uniaxial stress then its intensity slightly decreases  (see Fig.\ref{E0E1b}). The increase of the energy gap from 4.27(1) meV to 4.48(1) meV is again similar to the results with hydrostatic pressure \cite{bourdarot:2010}. Using the single-mode approximation, a larger decrease of intensity is expected. However, contrary to the excitation at \textbf{Q$_0$}, the incommensurate excitation at \textbf{Q$_1$} does not vanish in the AF state: its energy gap E$_1$ just shifts to higher energy. Therefore as the HO-AF transition is first order, it is possible that we are measuring a mixture of E$_1$ in the HO state at an energy transfer of $\sim$4.5meV, with E$_1$ in the AF state at higher energy. This may explain why the intensity decreases less than expected. In the case of hydrostatic pressure, the energy gap E$_1$ jumps from $\simeq5$meV to $\simeq8$meV \cite{bourdarot:2010}, but it may be smaller in the uniaxial stress case. Nevertheless, the initial evolutions of the gaps under uniaxial stress or hydrostatic pressure below the stress or pressure threshold in the HO state, are very similar and give the same critical energy gap E$_{0-\mathrm{crit}}\sim$1.2 meV.

\subsection{Thermal expansion and distribution of the lattice parameters a and c}
\label{IIID}
\begin{figure}
\begin{center}
	\includegraphics[width=.5\textwidth]{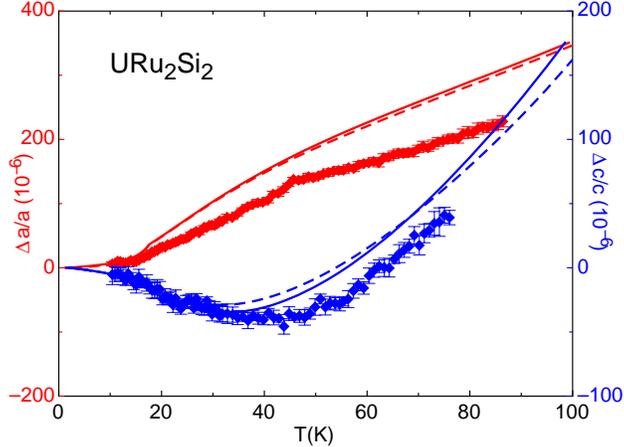}\\
\end{center}
\caption{Thermal expansions of \textbf{a} and \textbf{c} lattice parameters in \urs. Dots, full lines and dashed lines represent our NLD data and the results from dilatometry measurements taken from [\onlinecite{Devisser:1986}] and [\onlinecite{Hardy:2011}], respectively: red and blue colors are for the \textbf{a} and \textbf{c} lattice parameters.}
\label{fig:NLD_2}
\end{figure}

Using NLD, we have measured the temperature variation of the \textbf{a} and \textbf{c} lattice parameters in \urs\ at ambient pressure. Thermal expansions from 10K to $\approx$ 85K along the two axis have been inferred from the (2,0,0) and (0,0,4) Bragg reflections respectively. They were performed with increasing temperature. %The results below 10 K are not presented because the stick where the sample is fixed or the cryostat is moving for temperatures lower than T$\sim$5K (we suspect a problem of condensation of helium in the cryostat tail). As the sample must not move during the measurement, these data were removed.
The thermal expansion value at 10 K given by dilatometry\cite{Devisser:1986} was used to normalize our data. The results are shown on Fig.\ref{fig:NLD_2} and compared to a dataset obtained by means of a three-terminal capacitance method (from [\onlinecite{Devisser:1986,Hardy:2011}]). Along the \textbf{a}-axis, we observe two clear changes in the slope at T$_{0} \sim 17.5$ K and $T_{\chi} \sim 45$ K. The first one corresponds to the transition from HO to the large moment AF phase. The second one is located at the position of the maximum in the easy-axis magnetic susceptibility\cite{Palstra:1985}. Along \textbf{c}, T$_{0}$ is marked by an inflection point in the curve whereas T$_{\chi}$ corresponds to a local minimum. The two diffraction and bulk measurement datasets are only qualitatively consistent: the main features are seen but absolute values differ. We have no clear explanation for this discrepancy, which might come from a different sample thermalization.

Another consequence of the linear relation between the Larmor phase $\varphi$ and the lattice parameter $d$ is that the final beam polarization, $\mathcal{P}_{f}(\varphi_0)$, yields the cosine Fourier transform of the lattice spacing distribution function $f(d / d_0)$ through :
\begin{equation}
	\mathcal{P}_{f}(\varphi_0) = \mathcal{P}_{i} \cdot \exp \left( - \frac{\varphi_0^2 (\delta d/d_0)^2}{16 \ln 2} \right) \ ,
\label{eq:pol_phi}
\end{equation}
where $f(d / d_0)$ is assumed to be gaussian, $\mathcal{P}_{i}$ is the beam polarization extrapolated to zero phase, $\varphi_0$ is the total phase corresponding to the mean neutron wavelength ($\varphi_0\propto d_0 \propto \lambda$) and $\delta d/d_0$ is the FWHM of $f( d / d_0)$. In practice, we find that $\mathcal{P}_{i} \sim 0.72$ while the natural polarization provided by reflection on the Heusler monochromator and analyzer is $\sim 0.90$. The difference between usually natural polarization and $\mathcal{P}_{i}$ is due to the divergence of the neutron beam through RF coils.

\begin{figure}
\begin{center}
	\includegraphics[width=.5\textwidth]{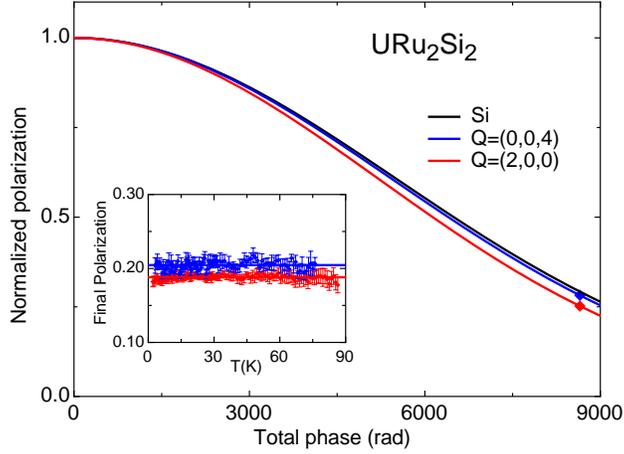}\\
\end{center}
\caption{Final beam normalized polarization as a function of the total phase $\varphi_0$. Black line represents the sake of comparison, a resolution curve which has been obtained by scattering on a ''perfect'' Si single crystal. Red and blue points correspond to the value of the normalized polarization measured at $\varphi_0 = 8650$ radians and T $\sim$ 2 K. Red and blue lines correspond to a law of the form of Eq. \ref{eq:pol_phi} applied to the respective cases of \textbf{a} and \textbf{c} lattice parameter. {\bf Inset:} Final beam polarization at $\varphi_0= 8650$ rad as a function of sample temperature.}
\label{fig:NLD_3}
\end{figure}

Fig.\ref{fig:NLD_3} shows the remaining normalized polarization ($\mathcal{P}_{f}(\varphi_0)/\mathcal{P}_{i}$) for $\varphi_0$=8650 rad at T = 2 K. This polarization gives a relative distribution width $\delta a/a_0 = 4.5(3)\cdot10^{-4}$ and $\delta c/c_0 = 4.4(2)\cdot10^{-4}$. A recent published experiment using NLD\cite{Niklowitz:2010}, has determined that the distribution of the lattice parameter \textbf{c} was surprisingly two times smaller than the distribution of the in-plane parameters \textbf{a}, {\it i.e.} $2.1 \cdot 10^{-4}$ and $4.05 \cdot 10^{-4}$ respectively. The outcome of our study is that both distributions are isotropic, independent of the temperature between 2 and $\sim$ 80 K, within error bars (see inset Fig. \ref{fig:NLD_3}). We would like to point out that those values have to be taken as high limits because extrinsic effects may slightly depolarized the beam and that the quality of our sample of \urs\ is as good as good as perfect high quality silicon single crystal (shown comparison on Fig.\ref{fig:NLD_3}). Nevertheless, the values of lattice parameter distributions are the same along \textbf{a} and \textbf{c}.

\section{Discussion}
\label{IV}
Within the linear elastic deformation regime, the hydrostatic pressure and uniaxial stresses are coupled with the strains by the well-known stress tensor which can be represented by the $c_{ij}$ matrix\cite{TI-Vol-D}. The pressure variation of the strains of all the U-U distances in the crystallographic cell as well as the ratio $\eta$=c/a were calculated using the elastic constants obtained with the ultrasonic-sound velocity extrapolated to T = 0K\cite{Wolf:1994}. Table \ref{strain} summarizes the derivative coefficients which are relevant for our study.
\begin{table}
\begin{tabular}{|c|c|c|}
\hline
X & p &$ \sigma^a$\\
\hline
\hline
stress  & (-p,-p,-p,0,0,0) & (0,-$\sigma$,0,0,0,0) \\
\hline
Symmetry& (a,a,c,$\frac{\pi}{2}$,$\frac{\pi}{2}$,$\frac{\pi}{2}$) & (a$_h$a$_v$,c,$\frac{\pi}{2}$,$\frac{\pi}{2}$,$\frac{\pi}{2}$) \\
&$I4/mmm$& $Immm$ \\
\hline
$\frac{\partial\widehat{\eta_h}}{\partial X}$  &  $\frac{c_{13}+c_{33}-c_{11}-c_{12}}{-2c_{13}^2+c_{33}(c_{11}+c_{12})}$ & $\frac{c_{13}(c_{13}+c_{11}-c_{12})-c_{12}c_{33}}{(c_{11}-c_{12})(c_{33}(c_{11}+c_{12})-2c_{13}^2)}$  \\
10$^{-3}$GPa$^{-1}$&1.2& 0.6 \\
\hline
$\frac{\partial\widehat{\eta_v}}{\partial X}$  &  $\frac{c_{13}+c_{33}-c_{11}-c_{12}}{-2c_{13}^2+c_{33}(c_{11}+c_{12})}$ & $\frac{c_{13}(c_{11}-c_{12}-c_{13})+c_{11}c_{33}}{(c_{11}-c_{12})(c_{33}(c_{11}+c_{12})-2c_{13}^2)}$ \\
10$^{-3}$GPa$^{-1}$&1.2& 5.4 \\
\hline
$\frac{\partial\widehat{\mathrm{a}_v}}{\partial X}$  & $\frac{c_{13}-c_{33}}{-2c_{13}^2+c_{33}(c_{11}+c_{12})}$ & $\frac{c_{13}^2-c_{12}c_{33}}{(c_{11}-c_{12})(c_{33}(c_{11}+c_{12})-2c_{13}^2)}$ \\
10$^{-3}$GPa$^{-1}$&-2.8 & -4.4 \\
\hline
\end{tabular}
\caption{Derivatives of the strain in the linear elastic regime of a$_v$,  $\eta_h$ and $\eta_v$ versus X with X=\{p,$\sigma^a$\}. The strain is defined as the relative variation of a parameter. It is represented with a hat on the parameter ($\widehat\zeta=\Delta\zeta/\zeta$=$\frac{\zeta(X)-\zeta(0)}{\zeta(0)}$).}
\label{strain}
\end{table}

Before going further, Table \ref{strain} can be only used in the case of hydrostatic pressure (column 2) or uniaxial stress along $\mathbf{ a}_v$ in the free lateral expansion conditions (column 3), as emphasized in Ref.[\onlinecite{Wei:2009}] (study made in the case of isotropic crystal). The free lateral expansion condition is fulfilled only if the ratio $\kappa\geqslant2$. For our small crystal or any crystal with $\kappa<$ 2, Table \ref{strain} cannot be used. In the case of very thin samples with no free lateral expansion, the set-up fulfills the experimental conditions for a uniaxial strain where $\frac{\partial\widehat{\mathrm{a}_v}}{\partial \sigma^a}$ can be estimated to $\sim -1/c_{11}=-3.9$ GPa$^{-1}$. In this condition, the critical stress for \urs will increase by 12\%. However as the experimental conditions are not under control if $\kappa\ll2$, the results are largely uncertain. In the experiments of Ref.[\onlinecite{Yokoyama:2005b}], as $\kappa \simeq 0.2$ their results cannot bring any reliable output for the critical stress.

Let us now compare the hydrostatic critical pressure $p_{x}$ generally found in the range of 0.5-0.7 GPa to the uniaxial critical stress along \textbf{a}-axis $\sigma_x^a\simeq$0.33 GPa. At the HO-AF transition, we consider that the relevant parameter ($\zeta$) is either the U-U distance (\textbf{a}) or the ratio $\eta$. At the  hydrostatic critical pressure or at the uniaxial critical stress, $\zeta$ should have the same critical value ($\zeta_{x}$). This is also true for $\widehat\zeta$  which is the relative variation of $\zeta$ ($\widehat\zeta=\Delta\zeta/\zeta$=$\frac{\zeta(X)-\zeta(0)}{\zeta(0)}$). So $\widehat{\zeta}_{x}=\frac{\partial\widehat{\zeta}}{\partial X}X_{x}$ is a constant independent of constraint $X$ (see figure in Table \ref{critical}) where X may be either hydrostatic pressure (p) or uniaxial stress ($\sigma^a$) ($X$=\{p or $\sigma^a$\}). To fullfill this condition, at the transition when \urs\ switches from the HO to the AF state, the experimental ratio $\frac{p_x}{\sigma_x^a}$ $\approx1.7\pm$0.3 should match the ratio $\frac{\partial \widehat{\zeta}}{\partial \sigma_a}/\frac{\partial \widehat{\zeta}}{\partial p}$ calculated in Table \ref{critical}.
Only $\mathbf{ a}_v$ fulfills this relation and the shortest U-U distance in the plane appears as the best candidate to control the magnetic properties of \urs. However to complete our discussion, the parameters $\eta_h$ and $\eta_v$ have been also considered in the following. In Table \ref{critical}, $\widehat{\zeta}_{x}$ have been calculated, at the critical hydrostatic pressure $p_{x}$ and at the critical uniaxial stress $\sigma_x^a$. It is interesting to note that with the shortest U-U distance in the plane as the relevant parameter, if the uniaxial stress is applied along the [1,1,0] direction, the critical stress $\sigma^{xx}_x$ should be larger ($\approx$ 0.8 GPa) than the hydrostatic critical pressure or uniaxial critical stress along \textbf{a}-axis. Also it is not possible to switch from HO state to AF state applying a stress along the \textbf{c}-axis in agreement with thermal expansion results.
\begin{table}
\begin{center}
\begin{tabular}{|c|c|c|c|c|c|}
\hline
   $\zeta$ & $\zeta_0$ & $\frac{\partial\widehat{\zeta}}{\partial \sigma^a}/\frac{\partial\widehat{\zeta}}{\partial p}$ & $ \widehat{\zeta}_{x}\ \mathrm{at}\ p_x$ & $\widehat{\zeta}_{x}\  \mathrm{at}\ \sigma_x^a$ & \multirow{4}{*}{\includegraphics[scale=0.215]{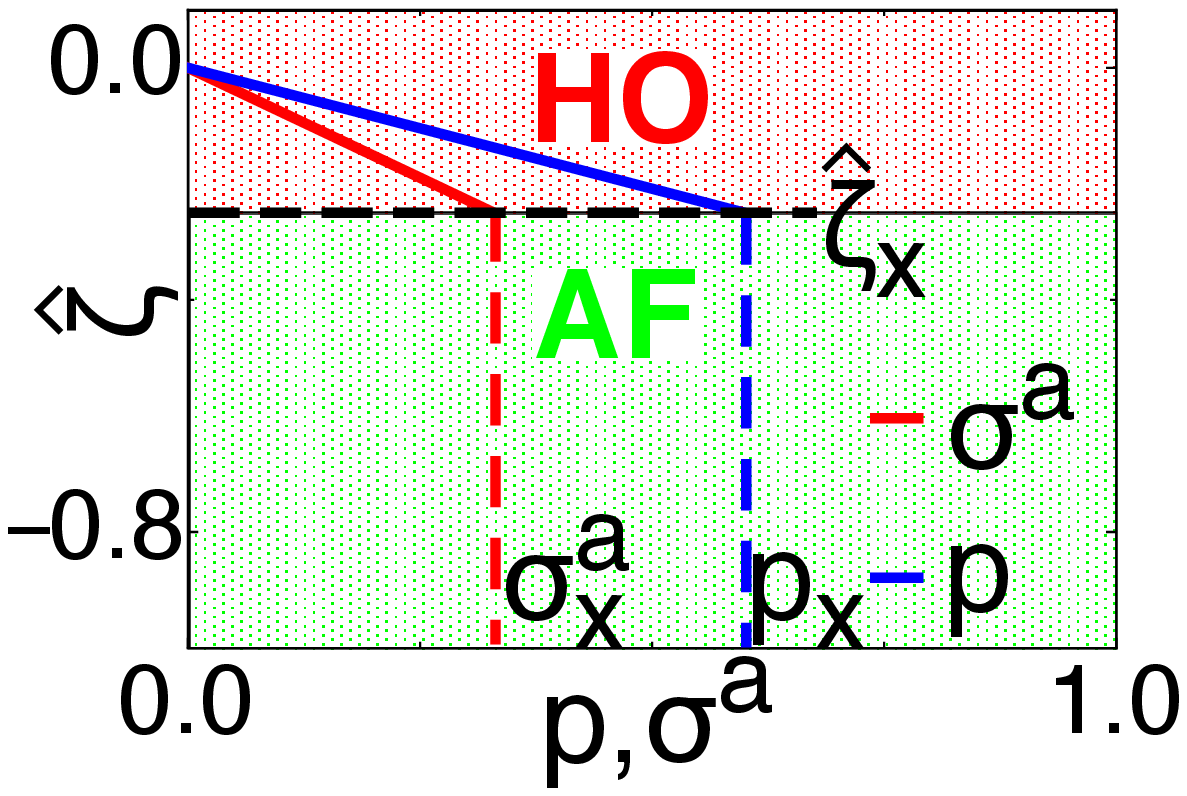}}\\
\cline{1-5}
  $\eta_h$ & 2.318 & 0.5 &  0.72$\cdot10^{-3}$ &  0.20$\cdot10^{-3}$&\\
  $\eta_v$ & 2.318 & 4.5 &  0.72$\cdot10^{-3}$ &  1.78$\cdot10^{-3}$&\\
   $a_v$ & 4.125\AA & 1.6 & -1.68$\cdot10^{-3}$ & -1.45$\cdot10^{-3}$ &\\
\hline
\end{tabular}
\end{center}
\caption{Table of the relevant parameters $\zeta$=$\eta_h$, $\eta_v$ and $a_v$: value at ambient pressure, ratio of the derivatives of $\widehat{\zeta}$ between uniaxial stress along \textbf{a}-axis and hydrostatic pressure, and values of $\widehat{\zeta}_{x}$ at the critical pressure p$_{x}$ =0.6 GPa and at $\sigma_x^a$=0.33 GPa. The figure on the right explains the assumption why  $\widehat{\zeta}_{x}$ should be equal at p$_x$ and $\sigma^a_x$, the critical value where \urs\ switches from HO to AF.} \label{critical}
\end{table}

The NLD result shows a difference between our distribution of the lattice parameter \textbf{c} and the previous distribution measurement  then there is almost no difference for the distributions of the lattice parameter \textbf{a}\cite{Niklowitz:2010}. Their \textbf{c} distribution is two times smaller than their \textbf{a} distribution, whereas we obtain an isotropic distribution along these two directions. Nevertheless the tiny AF moment does not reveal a large difference, namely 0.020(4) $\mu_B$ in our case to be compared to 0.012 $\mu_B$ in their case. Our larger distribution may be explained as we use a larger crystal $\simeq$ 100mm$^3$ to realize reliable inelastic experiments.

%Considering that the experiments are performed in a range of pressure that corresponds to the regime of linear elastic deformation.
According the phase diagram under pressure, an intrinsic AF moment exists only in the AF state (with a value m$_{AF}\simeq $ 0.36-0.40 $ \mu_B$ at low temperature\cite{Hassinger:2010b}). This assumption means, in the regime of linear elastic deformation, that the AF moment exists only above (or below) a critical value $\zeta_x$ of the relevant parameter. Thus it is possible to calculate the AF volume which corresponds to the integration of the distribution of the relevant parameter above (or below) to the critical value $\zeta_x$. In this same way, the variation under hydrostatic pressure or uniaxial stress of the AF volume can be calculated: the variation of $\zeta$, which corresponds to the average value of the distribution  is given by the formula:  $\zeta(X)=\zeta_0(1+\frac{\partial\widehat{\zeta}}{\partial X} X)$, with $X$=\{p or $\sigma^a$\}. With the assumption of a gaussian distribution of the relevant parameters $\zeta$, the AF volume is given by:

\begin{eqnarray}
\mathrm{V}_{AF}/\mathrm{V}_0 & = & \int_{\zeta_{x}}^{\infty}{\frac{2\sqrt{\ln{2}}}{\delta \zeta \sqrt{\pi}}e^{-(\frac{2\sqrt{\ln{2}}(\zeta-\zeta(X))}{\delta \zeta})^2}d \zeta}\nonumber \\
  & = & 1/2*\mathrm{erfc}\left(\frac{2\sqrt{\ln{2}}\frac{\partial\widehat{\zeta}}{\partial X}(X_{x}-X)}{\delta \zeta/\zeta_0}\right)
  \label{form}
\end{eqnarray}
where V$_0$ is the total volume of the sample, and $\zeta_0$ is the value of the relevant parameter $\zeta$ at ambient pressure. The erfc function is the complementary error function which can be found in the scientific library of Python\cite{python} (Numpy and Scipy).

Formula \ref{form} shows that the slope of AF volume versus the pressure or the uniaxial stress increases when width of the distribution $\partial\zeta$ is smaller but also is larger for uniaxial stress than for hydrostatic pressure as the derivative of the strain $\frac{\partial\widehat{\zeta}}{\partial X}$ is larger. This may explain why with samples of bad quality (large distribution $\delta\zeta$) the moment has a smaller slope and increases linearly with pressure.

\begin{figure} %
\includegraphics[width=80mm]{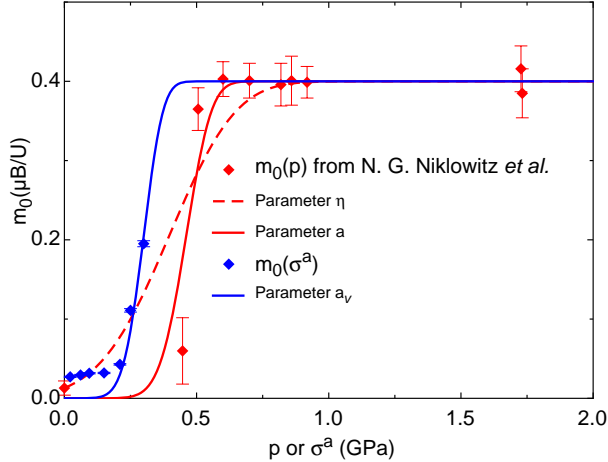}
\caption{Pressure and uniaxial stress dependences of m$_0$ with $\eta$ or a$_v$ as relevant parameters. The fits were made with m$_{AF}$=0.4$\mu_B$ and using the formula \ref{form}. The dashed and full lines correspond to $\eta$ and a$_v$ as relevant parameters respectively. For the hydrostatic pressure fit, the parameters were taken from Ref.[\onlinecite{Niklowitz:2010}], for uniaxial stress they come from this study. The derivatives $\frac{\partial\widehat{\zeta}}{\partial X}$ were taken from Table \ref{strain}.}
\label{meta}
\end{figure}

Fig.\ref{meta} represents the pressure and uniaxial stress variation of the AF moment (m$_{AF}*\sqrt{\mathrm{V}_{AF}(p)/\mathrm{V}_0}$) with $\mathbf{ a}_v$ or $\eta$ as relevant parameters. Our data under uniaxial stress along \textbf{a}-axis are compared to the hydrostatic pressure results of [\onlinecite{Niklowitz:2010}]. It is clear that $\eta$ does not appear to be the relevant parameter. With $\mathbf{ a}_v$ as relevant  parameter, the shape of the transition at $\sigma_x^a$ or p$_x$ is well explained. However, it is not possible to explain anymore the tiny moment at ambient pressure with only a simple gaussian distribution. We manage to reproduce the pressure dependence of the tiny AF moment with two other sources: either with an additional contribution to the \textbf{a} distribution (a tail in the distribution due to defects) or to intrinsic exotic nanostructured defects related to the high sensitivity of \urs\ to pressure, uniaxial stress or magnetic field \cite{Matsuda:2011}.

\begin{figure} %
\includegraphics[width=80mm]{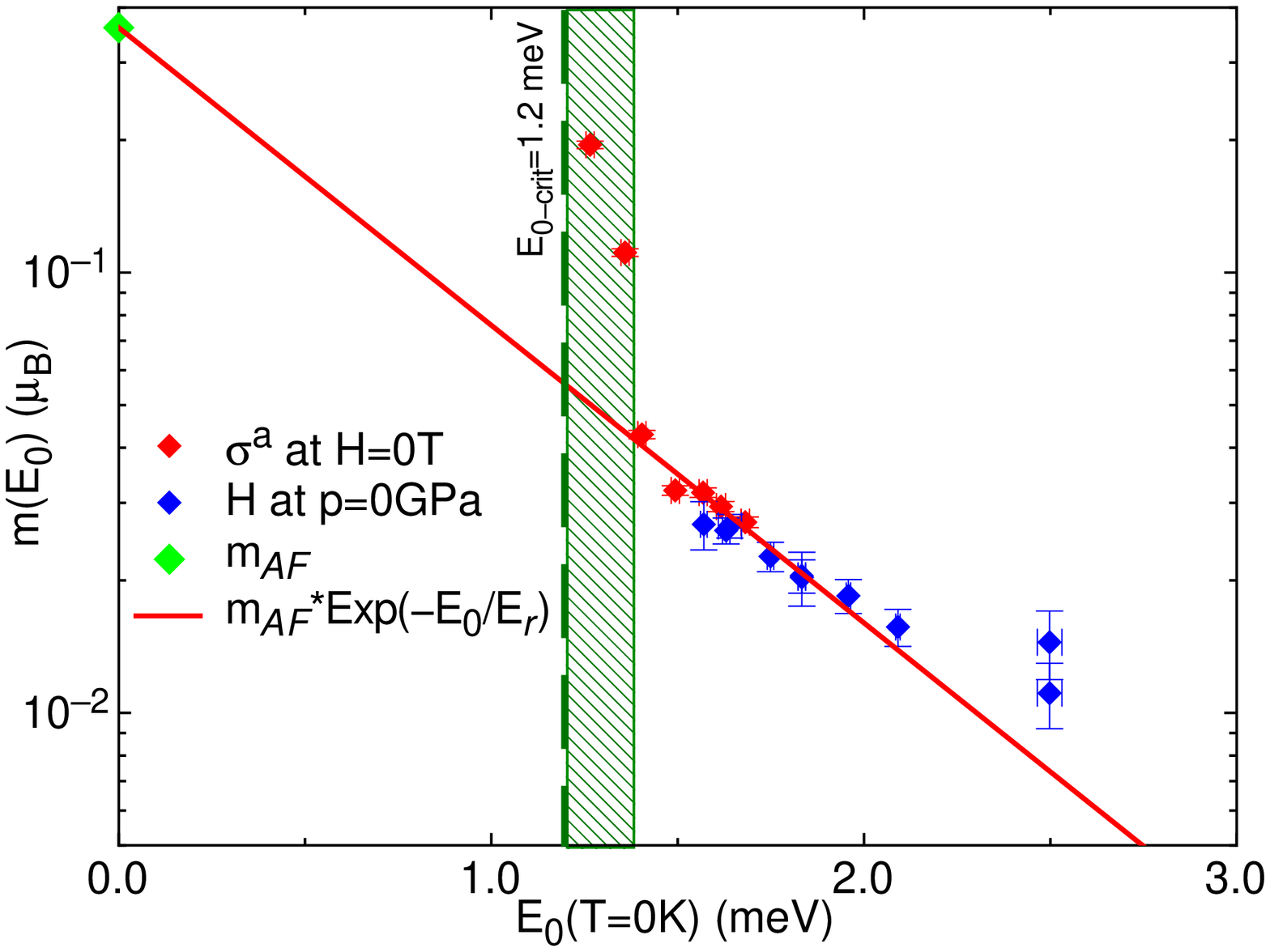}
\caption{Tiny AF moment at low temperature versus AF gap E$_0$ in a log scale from measurements under magnetic field \cite{Santini:2000,Bourdarot:2003,Bourdarot:2004} and this study under uniaxial stress. The green area is not taken into account in the fit because it corresponds to a mixture of both phases (HO and AF). The red curve corresponds to a fit with E$_r$= 0.64 meV and m$_{AF}$ = 0.36 $\mu_B$.}
\label{m0gap}
\end{figure}

The suspicion that these exotic defects come from the unusual effects is due to the fact that the detected tiny AF moment appears related to energy gap at \textbf{Q$_0$}, which is a signature of the hidden order\cite{Villaume:2008}, and to the AF moment m$_{AF}$ by the relation: m$_0(X)$=m$_{AF}$ e$^{-\frac{\mathrm{E}_0(\mathrm{X})}{\mathrm{E}_r}}$ with $\mathrm{E}_r\simeq0.64$ meV and X corresponding either to magnetic field studies at p=0 or uniaxial stress at H=0 (Fig.\ref{m0gap}). The hydrostatic pressure dependence of the AF energy gap E$_0$ was previously determined at low temperature \cite{bourdarot:2010}:  the energy gap E$_0(p)$ decreases linearly with pressure up to the critical pressure with a critical gap E$_{0-\mathrm{crit}}\simeq$ 1.1(1) meV. In this study, the uniaxial stress dependence of the AF energy gap E$_0$ has the same behavior with a larger slope but a smaller critical stress $\sigma_x^a$ leading to the same critical gap E$_{0-\mathrm{crit}}\simeq$ 1.2(1) meV.

A first order transition at $\sigma_x^a$ or p$_x$ indicates usually a strong repulsion between the two order parameters governing each side of this transition line where both order parameters can be mixed only if a coupling exists between them and in particular if both break time reversal symmetry. The invariance of the nuclear crystallographic structure between the PM and HO states with the loss of symmetric elements in the HO state keeping the atomic position at ambient pressure as  well as with hydrostatic pressure and also under uniaxial stress along \textbf{a}, is in agreement with the idea developed in Ref.[\onlinecite{Harima:2010,Kusunose:2011}] with the proposals for quadrupole or hexadecapole solutions for the hidden order. The ordering of any even-parity multipole will not break time reversal symmetry and thus at low stress mixing between HO and AF will not occur. However two resonant x-ray scattering measurements ruled out the possibility of any quadrupole ordering by resonant x-ray scattering\cite{Amitsuka:2010,walker:2011}. A hexadecapole remains a sound solution. Such a ground state was proposed in a  model based on a unified complex order parameter where the real part is a hexadecapole (not breaking time reversal symmetry) and the imaginary the magnetic dipole (breaking time reversal symmetry)\cite{Haule:2009,Kusunose:2011}. However, it was proposed that the critical stress along \textbf{a} (or [1,1,0]) should be two times smaller than the hydrostatic critical pressure\cite{Haule:2010}. It would be interesting to probe the critical stress along [1,1,0] direction, as we expect here a large increase of the critical stress compatible with the shortest U-U distance as the relevant parameter ($\simeq0.8$ GPa). 

On the other hand, mixture of HO state with AF state is possible in models with odd multipole ordering. The octupole model developed by Kiss and Fazekas\cite{Kiss:2005} seems in disagreement with the results of this paper as they have considered a coexistence of both order parameters under uniaxial stress. The dotriacontapole also breaks time reversal symmetry \cite{Cricchio:2009}. Both order parameter, HO and $m_z$ belong to the same group representation and then may generally be mixed. However according to the Ref.[\onlinecite{Cricchio:2009}], the tiny AF moment is not a real AF moment as proved by the latest NMR results \cite{Takagi:2007} where neither the large (0.3$\mu_B$) nor the small (0.02-0.04$\mu_B$) moments were detected, excluding a homogenous small static classical AF moment in \urs. Then the small signal measured by neutron scattering is the noncollinear dotriacontapole itself which has an extremely short-range stray field. However, such a multipole may only have a finite cross-section at high momentum transfer \cite{walker:2011} which seems not compatible with the small momentum transfer of the tiny moment. Moreover, there is no experimental trace of the dotriacontapole in the AF state as proposed in Ref.[\onlinecite{Cricchio:2009}]. Other contradiction: if the U-U distance is the relevant parameter, then the critical value for \textbf{a} occurs for a variation of 1.5\%, one order of magnitude larger than the present experimental results (see Table \ref{critical}).

In previous considerations, the 5$f$ localized character of the U atoms plays a key role. Other possibilities exist for the hidden order parameter corresponding to m(\textbf{Q$_0$}) based on more itinerant models such as a dynamic order parameter (symmetry breaking by dynamical antiferromagnetic fluctuations of the hidden order) \cite{Bernhoeft:2003,Elgazzar:2009,Oppeneer:2010}. In this case, the HO breaks time-reversal symmetry only for a short time. The theoretical Fermi Surface computed for the symmetry-broken state suggests that the body-centered translation vector is broken in the HO phase, in good agreement with the observation and the fact that the Fermi Surface of HO and AF are quite similar, $i.e.$ \textbf{Q$_{HO}$}=\textbf{Q$_{AF}$}, as seen in the experimental results\cite{Villaume:2008,Hassinger:2008}. Calculation of Fermi Surface in the orthorhombic structure symmetry, using the cell parameters induced by uniaxial stress, may ascertain this itinerant model.

\section{Conclusion}
\label{V}
Comparing the uniaxial stress along \textbf{a}-axis to hydrostatic pressure measurements, both phase diagrams are quite equivalent with a critical pressure almost two times smaller in the case of the stress (0.33 GPa compare to 0.6 GPa). The magnetic properties of \urs\ appear to be governed by the shortest  U-U distance (\textbf{a} parameter) and not by the ratio $\eta=c/a$. NLD results invalidate the simple model of large lattice parameter distribution to explain the tiny AF moment as extrinsic. The study of the hidden order state (more exactly of its fingerprint, the excitation E$_0$) under uniaxial stress indicates that this order can exist in a four-fold axis local symmetry as well in a two-fold axis symmetry. It is in agreement with a loss of the four-fold axis symmetry on the U site when entering in the HO state as for the group $P4_2/mnm$ (This space group is one of two possible space groups proposed by H. Harima\cite{Harima:2010}, but the only one which loses the four-fold axis on the U site).

The most promising model is the hexadecapole model developed in Ref.[\onlinecite{Haule:2009,Haule:2010}] and further discussed using symmetry argument in Ref.[{\onlinecite{Kusunose:2011}]. It will be interesting with this model to estimate the evolutions of the excitations at \textbf{Q$_0$} and \textbf{Q$_1$} and to compare them to experimental results already measured under pressure, stress and magnetic field\cite{Santini:2000,Bourdarot:2003,Bourdarot:2004}.

%The intriguing problem is the persistence of the tiny ordered moment in the low pressure phase (p$<$p$_x$) and of the superconductivity pocket in the high pressure phase (p$>$p$_x$). Keeping the idea that they can originate for particular defects, it is necessary to assume that long tails exist in the distribution of the pressure with an extension comparable to $\pm$ p$_x$ \cite{Matsuda:2011}. This may not correspond to a large lattice distribution in the correlated system as the electronic properties may change locally  \textcolor{magenta}{\sout{to realize} and keep the homogeneity of the lattice}. Microscopic local probe will certainly clarify if an unconventional intrinsic array of defects occur in \urs.

\section{Acknowledgements}
We acknowledge the financial support of the French Agence Nationale de la Recherche within the programs DELICE, CORMAT, SINUS and the european finance support ERC starting grant (NewHeavyFermion). We thank E. Hassinger D. Braithwaite and L. Malone for useful discussions.

\end{document}